\documentclass[twocolumn,showpacs,preprintnumbers,amsmath,amssymb]{revtex4}

\begin{document}
\title{Stochastic process leading to wave equations in dimensions higher than one}
\author{A.V. Plyukhin}
\email{aplyukhin@anselm.edu}
 \affiliation{ Department of Mathematics,
Saint Anselm College, Manchester, New Hampshire 03102, USA 
}

\date{\today}

\begin{abstract}
Stochastic processes  
are proposed whose master equations coincide  with
classical wave, telegraph, and Klein-Gordon equations. 
Similar to predecessors based on  
the Goldstein-Kac telegraph process, the model  describes 
the motion of particles with constant speed and  
transitions between discreet allowed velocity directions.
A new ingredient is that transitions into a given velocity state 
depend on spatial derivatives of other states populations,
rather than on populations themselves. This feature requires the sacrifice of
the single-particle character of the model, 
but allows to imitate the Huygens' principle and to recover wave equations 
in arbitrary dimensions.

\end{abstract}


\maketitle
\section{Introduction}
Can a wave, which is in general an essentially dynamical process, 
be mapped onto a
kinematic stochastic model like random walks?
This question has a long history and was approached from many different
perspectives, ranging from pragmatic  
(numerical simulation of wave-related phenomena)  
to fundamental (interpretation of
quantum mechanics) with   many important incentives in between, such as 
effects of inertia in heat transfer, light propagation in turbid media,
turbulence diffusion, etc.
While random walks are natural underlying processes for parabolic equations 
of diffusion type, the connections between stochastic motion and hyperbolic 
wave equations 
are less obvious and often restricted to one spatial dimension ($1D$). 
Perhaps the best-known example is
the telegraph equation
$f_{tt}+\frac{1}{\tau}\, f_t=c^2\Delta f$,
which describes propagation of waves in media with
losses (with characteristic dissipation time $\tau$). 
In $1D$  the equation can be readily derived from  
persistent random walk with constant speed and Poissonian velocity reversals
(often referred to as the Goldstein-Kac process)~\cite{GK,Weiss,Kolck}, but 
the same walk extended to higher dimensions does not evolve according to the 
telegraph equation~\cite{Weiss,Bartlett,Godoy,Masoliver,Masoliver2}. 
This feature is  generic and  
inherited  in  many related problems, in particular
of mapping  relativistic quantum wave 
equations onto classical random walks~\cite{Gaveau,Ord,PS}. 
Such mappings are 
typically designed in $1D$, and  the
extension to higher spatial dimensions requires the formal replacement of
the space-variable derivative by the 
gradient,  $\partial/\partial x \to \nabla$. 
It was noted by many authors that this approach may be  
inconsistent  since in general
it is impossible to construct the random walk with desirable properties
(governed by a master equation of desirable form) as a mere superposition of 
independent one-dimensional walks.  

As will be discussed below, 
the difficulty of extending wave-particle isomorphism 
beyond $1D$
is not related to stochastic nature of the 
random walk models, but rather 
originates from the inability for 
a {\it single-particle} motion, neither stochastic nor deterministic, 
to imitate 
the Huygens' principle in dimensions higher than one~\cite{remark}.
On the other hand, 
there are often no reasons  to 
restrict oneself to single-particle models, which cannot reproduce
physically relevant negative  solutions anyway.
In this paper we discuss a model {\it \`a la} Goldstein-Kac based on 
an ensemble of classical particles moving with a fixed speed and subjected 
to transitions between discreet allowed directions of motion.
It is shown  that transitions can be chosen in a form 
which allows to
recover wave-like  equations for the ensemble's distribution function 
in any dimension.

\section{Wave equation}
As well known, the classical wave equation in $1D$
$f_{tt}=c^2 f_{xx}$
can be mapped onto a totally deterministic kinematic model.
Thanks to the factorization $(\partial/\partial t-c\partial/\partial x)
(\partial/\partial t+c\partial/\partial x)f=0$, 
the general solution can be written
as the superposition  of  the functions $f^\pm(x,t)$, 
which satisfy the equations 
\begin{eqnarray}
\frac{\partial f^+}{\partial t}=-c\,\frac{\partial f^+}{\partial x},\qquad
\frac{\partial f^-}{\partial t}=c\,\frac{\partial f^-}{\partial x},
\label{1}
\end{eqnarray}
and can be interpreted as 
the distribution functions for independent particles, or for 
an ensemble of single particles, moving freely with constant 
speed $c$ 
in positive and negative directions, respectively.
Eqs. (\ref{1})  also can be written in terms 
the total distribution function $f=f^++f^-$ and the current
$J=c\,(f^+-f^-)$,
\begin{eqnarray}
\frac{\partial f}{\partial t}=-\frac{\partial J}{\partial x},\qquad
\frac{\partial J}{\partial t}=-c^2\,\frac{\partial f}{\partial x}.
\label{2}
\end{eqnarray}
One might suggest that for higher dimensions 
the proper generalization of Eqs.(\ref{2}) should read
\begin{eqnarray}
\frac{\partial f}{\partial t}=-\vec\nabla\cdot\vec J,\qquad
\frac{\partial \vec J}{\partial t}=-c^2\,\vec\nabla f,
\label{3}
\end{eqnarray}
which indeed immediately gives the multidimensional wave equation 
for $f$, 
\begin{eqnarray}
\frac{\partial^2 f}{\partial t^2}=
c^2\,\Delta f,
\label{WE3}
\end{eqnarray}
as well as the conservation law for the current vorticity  
\begin{eqnarray}
\frac{\partial}{\partial t}\,  \vec\nabla\times\vec J=0.
\end{eqnarray} 
However,  it is easy to see that 
the equations (\ref{3}) do not follow from multidimensional  
generalization  of Eqs. (\ref{1}) and therefore cannot describe 
a single-particle motion in $D>1$.

It is instructive to illustrate the last statement explicitly for two 
spatial dimensions ($2D$), 
assuming that 
particles move with a constant speed $c$ and can be in one of four 
velocity states $(x,\pm)$, 
$(y,\pm)$, corresponding to motions along positive ($+$) or 
negative ($-$)
directions of the two Cartesian axes. 
Let $f_x^\pm(x,y,t)$ and $f_y^\pm(x,y,t)$  
be the corresponding distribution functions. 
(From here on, we use subscripts $\alpha=x,y,z$ only 
to refer to
vector components, not to partial derivatives).   
For freely moving particles there are no transitions between the states,
so the two-dimensional version of Eqs. (\ref{1}) reads  as follows: 
\begin{eqnarray}
\frac{\partial f_x^\pm}{\partial t}=
\mp c\,\frac{\partial f_x^\pm}{\partial x},\qquad
\frac{\partial f_y^\pm}{\partial t}=\mp c\,\frac{\partial f_y^\pm}{\partial y}.
\label{4}
\end{eqnarray}
Let $f_x=f_x^++f_x^-$, $f_y=f_y^++f_y^-$, and 
$J_x(x,y,t)=c(f_x^+-f_x^-)$ and $J_y(x,y,t)=c(f_y^+-f_y^-)$ are Cartesian 
components of the current $\vec J=\vec i J_x+\vec j J_y$. 
Adding and subtracting  Eqs. (\ref{4}) one obtains
\begin{eqnarray}
\frac{\partial f_x}{\partial t}=-\frac{\partial J_x}{\partial x},\qquad
\frac{\partial J_x}{\partial t}=-c^2\,\frac{\partial f_x}{\partial x}
\end{eqnarray}
for particles moving along $x$-axis, and 
\begin{eqnarray}
\frac{\partial f_y}{\partial t}=-\frac{\partial J_y}{\partial y},\qquad
\frac{\partial J_y}{\partial t}=-c^2\,\frac{\partial f_y}{\partial y},
\end{eqnarray}
for particles moving  along $y$-axis. These equations 
do not form the  closed system (\ref{3}) for   
the total distribution 
$f(x,y,t)=f_x+f_y$ and the current $\vec J$,
\begin{eqnarray}
\frac{\partial f}{\partial t}&=&-\vec \nabla\cdot\vec J,\\
\frac{\partial \vec J}{\partial t}&=&
-c^2\left\{\vec i \,\frac{\partial f_x}{\partial x}+
\vec j \,\frac{\partial f_y}{\partial y}\right\}\ne
-c^2\vec \nabla f.\nonumber
\end{eqnarray}
Therefore the  two-dimensional wave equation  is not recovered.

Let us now generalize Eqs. (\ref{4}) by allowing  
transitions between $(x,+)$ and 
$(x,-)$ states with transition rate $g_x(x,y,t)$,  
and between $(y,+)$ and $(y,-)$ states with transition rate $g_y(x,y,t)$,
\begin{eqnarray}
\frac{\partial f_x^\pm}{\partial t}=
\mp c\,\frac{\partial f_x^\pm}{\partial x}\mp g_{x},\qquad
\frac{\partial f_y^\pm}{\partial t}=
\mp c\,\frac{\partial f_y^\pm}{\partial y}\mp g_{y}.
\label{gen}
\end{eqnarray}
(Note that no transitions are still allowed between $x$ and $y$ states.)
The presence of transitions does not violate the conservation of 
the number of particles, 
so the continuity
equation $ \partial f/\partial t=-\vec \nabla\cdot\vec J$ still holds, 
while the equation
for  the current  now reads
\begin{eqnarray}
\label{mark}
\frac{\partial \vec J}{\partial t}=
-c^2\left\{\vec i \,\left(\frac{\partial f_x}{\partial x}+\frac{2}{c}g_x\right)+
\vec j \,\left(\frac{\partial f_y}{\partial y}+\frac{2}{c}g_y\right)\right\}.
\label{pig}
\end{eqnarray}
By choosing transition rates  in the form 
\begin{eqnarray}
g_x=\frac{c}{2}\, \frac{\partial f_y}{\partial x},\qquad 
g_y=\frac{c}{2}\, \frac{\partial f_x}{\partial y}
\label{rates}
\end{eqnarray}
the right-hand side of Eq.(\ref{pig}) is completed  
to the gradient,
$\partial \vec J/\partial t=-c^2\vec\nabla f$.
Thus the system (\ref{3}) is recovered, and therefore 
the total distribution $f$ is governed by the
two-dimensional wave equation (\ref{WE3}).

With transition rates (\ref{rates}), the partial motions along $x$ and $y$ axes
become statistically coupled even though there are no direct transitions
$(x,\pm)\leftrightarrow (y,\pm)$. Note that an equation for 
the particle distribution
in a given state, say $(x,+)$,  
\begin{eqnarray}
\frac{\partial f_x^+}{\partial t}=
-c\,\frac{\partial f_x^+}{\partial x}-
\frac{c}{2}\,\frac{\partial f_y}{\partial x},
\end{eqnarray}
has  a form of the 
conservation law $\partial f_x^+/\partial t=-\partial J_x^+/\partial x$, where
the flux $J_x^+=c\left(f_x^+ + \frac{1}{2}f_y\right)$ is 
determined not only by the density of 
the given state $f_x^+$ but also by local particles in other velocity
states $f_y^+$ and $f_y^-$. This is reminiscent of 
the Huygens' principle, according to which
every point of a wave front propagating with a speed $c$ is 
the source of secondary waves 
that spread out in all directions with the  same speed $c$.
The resemblance is achieved at the expense of
losing the single-particle status of the original $1D$ model.
It is easy to see that with transition rates (\ref{rates}), 
proportional to derivatives of state populations,   
the partial distributions
$f^\pm_x$ and $f^\pm_y$  are not positively defined. Then Eqs.(\ref{gen}) 
can not describe a single-particle motion, but
should  be interpreted 
as the equations   for  corresponding perturbations in   
an ensemble of particles.

The generalization of the scheme for $3D$ is straightforward
and reads
\begin{eqnarray}
\frac{\partial f_\alpha^\pm}{\partial t}=
\mp c\,\frac{\partial f_\alpha^\pm}{\partial \alpha}\mp g_{\alpha},\qquad
\alpha=x,y,z
\label{gen3}
\end{eqnarray}
with  
\begin{eqnarray}
g_x=\frac{c}{2} \left(
\frac{\partial f_y}{\partial x}+\frac{\partial f_z}{\partial x}
\right)
\label{rates2}
\end{eqnarray}
and similar expressions for $g_y$ and $g_z$. The equations 
(\ref{gen3}) and (\ref{rates2}) lead to the system (\ref{3}) and therefore
to the wave equation  in $3D$.

A trick of coupling of partial  motions with transition rates
in the form (\ref{rates}) or  (\ref{rates2}) is quite  generic
and can be applied to derive other multidimensional wave-like equations.
Two more examples are presented below.

\section{Telegraph equation}

A hybrid of the wave and diffusion equations, the telegraph equation 
has  a form 
\begin{eqnarray}
\frac{\partial^2 f}{\partial t^2}+
\frac{1}{\tau}\,\frac{\partial f}{\partial t} =c^2\Delta f
\label{TE3}
\end{eqnarray}
and 
describes  the propagation of waves 
in media with losses. It is also often used as an approximation 
to treat diffusion processes 
beyond the overdamped limit and in a turbulent 
medium~\cite{GK,Weiss,Broeck}.
A one-dimensional version of Eq.(\ref{TE3})  
can be derived as a master equation for the Goldstein-Kac 
stochastic process -  
a dichotomous  persistent random walk  in which 
a particle moves in $1D$  with the velocity fluctuating between $c$ and
$-c$. 
Using the same notations as in the previous section, 
let $f^+(x,t)$ and $f^-(x,t)$ be
the probability density for the particle moving to the right and to the 
left, respectively.  Reversals of velocity are Poisson distributed and  
occurring with the rate $1/2\tau$. The processes  is described by
equations
\begin{eqnarray}
\frac{\partial f^+}{\partial t}&=&-c\,\frac{\partial f^+}{\partial
  x}-\frac{1}{2\tau}\,(f^+-f^-),\label{8}\\
\frac{\partial f^-}{\partial t}&=&
+c\,\frac{\partial f^-}{\partial x}+\frac{1}{2\tau}\,(f^+-f^-).\nonumber
\label{9}
\end{eqnarray}
In terms of  the total distribution function  $f=f^++f^-$ and 
the current $J=c\,(f^+-f^-)$, the equations take the form
\begin{eqnarray}
\frac{\partial f}{\partial t}=-\frac{\partial J}{\partial x},\qquad
\frac{\partial J}{\partial t}+\frac{1}{\tau}J=-c^2\,\frac{\partial f}{\partial x}.
\label{10}
\end{eqnarray} 
This   leads immediately   
to the $1D$ telegraph equation for  $f$ 
\begin{eqnarray}
\frac{\partial^2 f}{\partial t^2}+
\frac{1}{\tau}\,\frac{\partial f}{\partial t}
=c^2\frac{\partial^2 f}{\partial x^2},
\label{7}
\end{eqnarray}
and also for $J$ and each 
components $f^+$ and $f^-$.

The isomorphism between persistent random
walk and dissipative wave propagation, 
though very attractive from many points of view,  
does not go  beyond $1D$. 
It is true that the multi-dimensional  telegraph equation (\ref{TE3})
follows readily from
the generalization of (\ref{9})
\begin{eqnarray}
\frac{\partial f}{\partial t}=-\vec\nabla\cdot \vec J,\qquad
\frac{\partial \vec J}{\partial t}+\frac{1}{\tau}\vec J=-c^2\vec\nabla f.
\label{hyp}
\end{eqnarray}
However,  these generalized equations do not follow
merely from multi-dimensional extension of 1D random walk (\ref{8}). 
For instance, such an extension for $2D$ has a form
\begin{eqnarray}
\frac{\partial f_x^+}{\partial t}&=&-c\,\frac{\partial f_x^+}{\partial x}
+\frac{1}{2\tau}\,(-3f_x^++f_x^-+f_y),
\label{origin}\\
\frac{\partial f_x^-}{\partial t}&=&+c\,\frac{\partial f_x^-}{\partial x}
+ \frac{1}{2\tau}\,(-3f_x^-+f_x^++f_y),\nonumber\\
\frac{\partial f_y^+}{\partial t}&=&-c\,\frac{\partial f_y^+}{\partial y}
+\frac{1}{2\tau}\,(-3f_y^++f_y^-+f_x),\nonumber\\
\frac{\partial f_y^-}{\partial t}&=&+c\,\frac{\partial f_y^-}{\partial y}
+ \frac{1}{2\tau}\,(-3f_y^-+f_y^++f_x),\nonumber
\end{eqnarray}
where  $f_x=f_x^++f_x^-$ and $f_y=f_y^++f_y^-$.
Adding and subtracting lead to the following equations for the 
partial densities
\begin{eqnarray}
\frac{\partial f_x}{\partial t}&=&-\frac{\partial J_x}{\partial x}
+\frac{1}{\tau}(-f_x+f_y) 
\label{aux1},\\
\frac{\partial f_y}{\partial t}&=&-\frac{\partial J_y}{\partial y}
+\frac{1}{\tau}(-f_y+f_x),\nonumber
\end{eqnarray}
and for the current components
\begin{eqnarray}
\frac{\partial J_x}{\partial t}&=&-c^2\,\frac{\partial f_x}{\partial x}
- \frac{1}{\tau}J_x, 
\label{aux2}\\
\frac{\partial J_y}{\partial t}&=&-c^2\,\frac{\partial f_y}{\partial y}
- \frac{1}{\tau}J_y.\nonumber
\end{eqnarray} 
Thus, for the total distribution function
$f=f_x+f_y$ and the current $\vec J=\vec i J_x+\vec j J_y$ 
one obtains
\begin{eqnarray}
\frac{\partial f}{\partial t}&=&-\vec\nabla\cdot\vec J,
\label{hyp2}
\\
\frac{\partial \vec J}{\partial t}+\frac{1}{\tau}\vec J&=&
-c^2\left\{\vec i \,\,\frac{\partial f_x}{\partial x}+
\vec j \,\,\frac{\partial f_y}{\partial y}\right\}\ne
-c^2\vec \nabla f.\nonumber
\end{eqnarray}
This differs from Eqs. (\ref{hyp}),  
and therefore the two-dimensional telegraph equation is not recovered.

Let us modify Eqs.(\ref{origin}) in precisely the same way as
in the previous section. Namely, in addition to transitions
with isotropic rate $1/2\tau$, let us 
introduce anisotropic transitions
$(x,+)\leftrightarrow(x,-)$ and $(y,+)\leftrightarrow (y,-)$ with the rates
$g_x$ and $g_y$, respectively,
\begin{eqnarray}
\frac{\partial f_x^+}{\partial t}&=&-c\,\frac{\partial f_x^+}{\partial x}
-g_x+\frac{1}{2\tau}\,(-3f_x^++f_x^-+f_y),
\label{TEsystem}\\
\frac{\partial f_x^-}{\partial t}&=&+c\,\frac{\partial f_x^-}{\partial x}
+g_x+ \frac{1}{2\tau}\,(-3f_x^-+f_x^++f_y),\nonumber\\
\frac{\partial f_y^+}{\partial t}&=&-c\,\frac{\partial f_y^+}{\partial y}
-g_y+\frac{1}{2\tau}\,(-3f_y^++f_y^-+f_x),\nonumber\\
\frac{\partial f_y^-}{\partial t}&=&+c\,\frac{\partial f_y^-}{\partial y}
+g_y+ \frac{1}{2\tau}\,(-3f_y^-+f_y^++f_x)\nonumber
\end{eqnarray}
with $g_\alpha$ given by Eq. (\ref{rates}), 
\begin{eqnarray}
g_x=\frac{c}{2}\, \frac{\partial f_y}{\partial x},\,\,\,\,\,\, 
g_y=\frac{c}{2}\, \frac{\partial f_x}{\partial y}.
\end{eqnarray}
In this case 
Eqs.(\ref{aux1}) for $\partial f_\alpha/\partial t$ 
do not change, 
while Eqs.(\ref{aux2}) for $\partial J_\alpha/\partial t$ take the form 
\begin{eqnarray}
\frac{\partial J_x}{\partial t}+\frac{1}{\tau}J_x&=&
-c^2\,\frac{\partial f}{\partial x}
, \\
\frac{\partial J_y}{\partial t} + \frac{1}{\tau}J_y&=&
-c^2\,\frac{\partial f}{\partial y}.\nonumber
\end{eqnarray} 
Thus both equations (\ref{hyp}) are satisfied, from where 
the two-dimensional telegraph equation (\ref{TE3}) for $f$  follows 
immediately.

The generalization of the system (\ref{TEsystem}) for $3D$ is straightforward,
i.e
the equation for $f_x^\pm$ takes the form
\begin{eqnarray}
\frac{\partial f_x^\pm}{\partial t}&=&\mp c\,\frac{\partial f_x^\pm}{\partial x}
\mp g_x+\frac{1}{2\tau}(-5f_x^\pm+f_x^\mp+f_y+f_z),
\nonumber
\end{eqnarray}
with $f_z=f_z^++f_z^-$ and $g_x$ given by Eq. (\ref{rates2}).

\section{Klein-Gordon equation}
The combination of a substitution $f= \exp(-t/2\tau)\psi$ and 
analytic continuation of the transition rate to the imaginary value
$1/2\tau\to -i\,\lambda$ converts the telegraph equation (\ref{TE3}) 
into the Klein-Gordon equation
\begin{eqnarray}
\frac{\partial^2 \psi}{\partial t^2}=c^2
\Delta\psi-\lambda^2\,\psi.
\label{KGE}
\end{eqnarray}  
This observation was exploited in many works 
discussing 
the possibility of the  connection between
random walks and quantum mechanics.
Yet
the concept of a complex transition rate does not look particularly attractive.

An alternative stochastic model  leading to the Klein-Gordon equation   
in $1D$ without recourse to analytic continuation
was discussed  in~\cite{PS}.
In this model transitions between two velocity states are governed by a
``guiding field'' 
$E(x,t)$ which is coupled to the total particle distribution $f(x,t)$ 
via the Poisson 
equation.  Using the same method  as in the previous sections, the model can
be extended for $2D$ as follows:                                                
\begin{eqnarray}
\frac{\partial f_x^+}{\partial t}&=&-c\,\frac{\partial f_x^+}{\partial x}
-g_x+\frac{1}{2} a E_x,
\label{cat1}\\
\frac{\partial f_x^-}{\partial t}&=&+c\,\frac{\partial f_x^-}{\partial x}
+g_x- \frac{1}{2} a E_x,\nonumber\\
\frac{\partial f_y^+}{\partial t}&=&-c\,\frac{\partial f_y^+}{\partial y}
-g_y+ \frac{1}{2} a E_y,\nonumber\\
\frac{\partial f_y^-}{\partial t}&=&+c\,\frac{\partial f_y^-}{\partial y}
+g_y-\frac{1}{2} a E_y.\nonumber
\end{eqnarray}
Here the transition rates 
$g_\alpha$ are given by Eq. (\ref{rates}), 
\begin{eqnarray}
g_x=\frac{c}{2}\, \frac{\partial f_y}{\partial x},\qquad
g_y=\frac{c}{2}\, \frac{\partial f_x}{\partial y},
\label{cat2}
\end{eqnarray} 
the field $\vec E=\vec i E_x+\vec j E_y$ satisfies the Poisson equation
\begin{eqnarray}
\vec\nabla\cdot\vec E=b f,
\label{PE}
\end{eqnarray}
$a$ and $b$ are constants.  
For the total distribution  
$f=f_x+f_y=f_x^++f_x^-+f_y^++f_y^-$
 and the current
$\vec J=c(f_x^+-f_x^-)\vec i+c(f_y^+-f_y^-)\vec j$ 
one obtains  from Eqs.(\ref{cat1}) 
\begin{eqnarray}
\frac{\partial f}{\partial t}&=&-\vec\nabla\cdot\vec J,
\label{q}\\
\frac{\partial \vec J}{\partial t}&=&-c^2\vec\nabla f+a c\vec E.\nonumber
\end{eqnarray}
Together with (\ref{PE}), these equations give the 
Klein-Gordon equation (\ref{KGE}) for $f$  
with $\lambda^2=a\,b\,c$. 
The possibility to interpret  
this approach in 
the spirit of the  Dirac sea model was discussed in~\cite{PS}.

The generalization of this scheme for  $3D$ is
straightforward and involves  transition rates $g_\alpha$ 
in the  form (\ref{rates2}). For instance, the equations for
$f_x^{\pm}$ take the form
\begin{eqnarray}
\frac{\partial f_x^+}{\partial t}&=&-c\,\frac{\partial f_x^+}{\partial x}
-\frac{c}{2}\,\left( \frac{\partial f_y}{\partial x}+\frac{\partial f_z}{\partial x}\right)
+\frac{1}{2} a E_x,\\
\frac{\partial f_x^-}{\partial t}&=&+c\,\frac{\partial f_x^-}{\partial x}
+\frac{c}{2}\,\left( \frac{\partial f_y}{\partial x}+\frac{\partial f_z}{\partial x}\right)
- \frac{1}{2} a E_x,\nonumber
\end{eqnarray} 
which gives
\begin{eqnarray}
\frac{\partial f_x}{\partial t}&=&-\frac{\partial J_x}{\partial x},\\
\frac{\partial J_x}{\partial t}&=&-c^2\,\frac{\partial f}{\partial x}+ a c E_x.
\nonumber
\end{eqnarray}
These and similar equations for $y$ and $z$ components
lead to (\ref{q}) and therefore to the Klein-Gordon equations in $3D$.
Note that if $\vec E$ is a conservative vector field, 
$\vec\nabla\times \vec E=0$, then it follows from (\ref{q}) that 
the vorticity of the current $\vec\nabla\times \vec J$ (``spin'') 
is a constant of motion.

\section{Conclusion} 

In this paper we constructed 
systems of partial differential equations of first order which 
can be interpreted in terms of stochastic motion of an ensemble of 
particles and lead to wave equations for the particle distribution 
in any dimensions.   
There is nothing unusual, of course, in wave-like dynamics in a system of
interacting particles.  
However, the presented model attempts to picture waves not as
a dynamical process but  kinematically, assuming that  
the particles move with speed equal to the phase 
speed of the wave and represent medium  excitations  
rather than medium constituents. 
Technically, the approach is 
based on a simple trick of 
``completing the gradient''.  Namely, 
an appropriate $1D$ single-particle process is extended to higher dimensions 
supplemented with additional terms $g_\alpha$, which make the equation for the 
local current to involve
the gradient of the total distribution (e.g.,
$\partial\vec J/\partial t=-c^2\vec\nabla f$ for the wave equation).
The additional terms $g_\alpha$
can be interpreted as transition rates between states
with opposite velocity directions
and make the model consistent with the Huygens' principle.
The discreteness  of velocity directions, assumed in this approach,  
may look physically artificial, yet technically it is essential.  
The model can be readily extended for any finite number of 
allowed directions, but not for a continuum of directions. 

\section{Acknowledgment}
I thank  Gregory Buck and Stephen Shea for discussion and
Bira van Kolck for correspondence.


\end{document}